\documentclass[12pt]{article}
\usepackage{comment}
\usepackage{hyperref}
\usepackage{enumitem}
\usepackage{amsmath,amssymb,epsf,cite,graphicx,subfigure}
\usepackage{chngcntr}
\usepackage{soul}
\counterwithout{equation}{section}
\newcommand{\beq}{\begin{equation}}
\newcommand{\eeq}{\end{equation}}

\def\<{\langle}
\def\>{\rangle}

\begin{document}
\baselineskip=15.5pt
\pagestyle{plain}
\setcounter{page}{1}

\begin{flushright}
{YITP-SB-14-10
\\
MIT-CTP-4541}
\end{flushright}

\vskip 0.8cm

\begin{center}
{\LARGE \bf Wormholes and entanglement in holography}
\vskip 0.5cm

Kristan Jensen$^{1,a}$ and Julian Sonner$^{2,3,b}$

\vspace{0.1cm}

{\it ${}^1$ C. N. Yang Institute for Theoretical Physics, SUNY \\ Stony Brook, NY 11794-3840, United States\\}

\vspace{.1cm}

{\it ${}^2$ Center for Theoretical Physics, Massachusetts Institute of Technology \\ Cambridge, MA 02139, United States\\}

\vspace{.1cm}

{\it ${}^3$ L.N.S., Massachusetts Institute of Technology \\ Cambridge, MA 02139, United States \\}

\vspace{0.1cm}

{\tt  ${}^a$kristanj@insti.physics.sunysb.edu, ${}^b$sonner@mit.edu \\}

\medskip

\end{center}

\vskip0.3cm

\begin{center}
{\bf Abstract}
\end{center}

In this essay, we consider highly entangled states in theories with a gravity dual, where the entangled degrees of freedom are causally disconnected from each other. Using the basic rules of holography, we argue that there is a non-traversable wormhole in the gravity dual whose geometry encodes the pattern of the entanglement. 

\bigskip

\begin{center}
Essay written for the Gravity Research Foundation \\  2014 Awards for Essays on Gravitation

\bigskip

\today

\end{center}
\newpage

\section*{Introduction} 

Consider two spatially separated spins, entangled into the state
\beq
\label{E:twoSpins}
|\Psi\>  = \frac{1}{\sqrt{2}}\left( |\uparrow\downarrow\> - |\downarrow\uparrow\>\right)\,,
\eeq
as in the Einstein-Podolsky-Rosen (EPR) thought experiment. The density matrix $\rho = |\Psi\>\<\Psi|$ is pure, but encodes non-trivial correlations. Single-spin measurements are uncorrelated, $\<\Psi| \vec{S}_{1,2}|\Psi\> = \vec{0}$, but the equal-time, connected correlator
\beq
\label{E:twoPoint}
G_{12}(\hat{n},\hat{m}) \equiv \< (\hat{n}\cdot \vec{S}_1) (\hat{m}\cdot \vec{S}_2)\> - \< \hat{n}\cdot\vec{S}_1\>\<\hat{m}\cdot \vec{S}_2\> = -\frac{ \hat{n}\cdot \hat{m}}{4}\,,
\eeq
is nonzero. $G_{12}$ signifies entanglement, insofar as it vanishes in product states. More precisely, $G_{12}\neq 0$ cannot, by causality, arise from interactions between the spins, and so indicates non-trivial correlations in $\rho$. Now, after preparing the state, separate the spins so that they are forever after out of causal contact. Then local interactions between the spins cannot decohere the correlations~\eqref{E:twoPoint} in~\eqref{E:twoSpins}, measured at later times. We term this sort of entanglement ``EPR entanglement.'' In a pure state, any connected correlator like $G_{12}$, with operators acting on causally disconnected degrees of freedom, provides a good measure of EPR entanglement. 

The AdS/CFT correspondence~\cite{Maldacena:1997re}, or holography, relates certain conformal field theories (CFTs) to quantum gravity on higher dimensional Anti-de-Sitter (AdS) spacetimes. Holography translates the physics of CFTs into statements about dual quantum gravity. Here we study the implications of EPR entanglement for AdS quantum gravity. It has been argued~\cite{Swingle:2009bg,VanRaamsdonk:2010pw} that the connectedness of the dual spacetime emerges from entanglement between spatially separated degrees of freedom. But how can a gravitational theory, hard-wired to respect causality, encode EPR correlations? In agreement with the ER-EPR conjecture~\cite{Maldacena:2013xja}, we find a simple and surprising answer: the gravity dual contains a wormhole, albeit a non-traversable one. The holographic dual of EPR entanglement is a wormhole.

\section*{From entanglement to wormholes}

We work with the duality between $\mathcal{N}=4$ $SU(N)$ super Yang-Mills (SYM) and IIB strings on AdS$_5\times\mathbb{S}^5$ when the string theory is approximately semiclassical IIB supergravity, which is the $N\gg \lambda \gg 1$ limit (for $\lambda$ the `t Hooft coupling). Setting the AdS radius to unity, the effective Newton constant on AdS$_5$ is $G_N \sim 1/N^2$, while the string length is $l_s \sim \lambda^{-1/4}$.

\begin{figure}
\begin{center}
	\includegraphics[width=7cm]{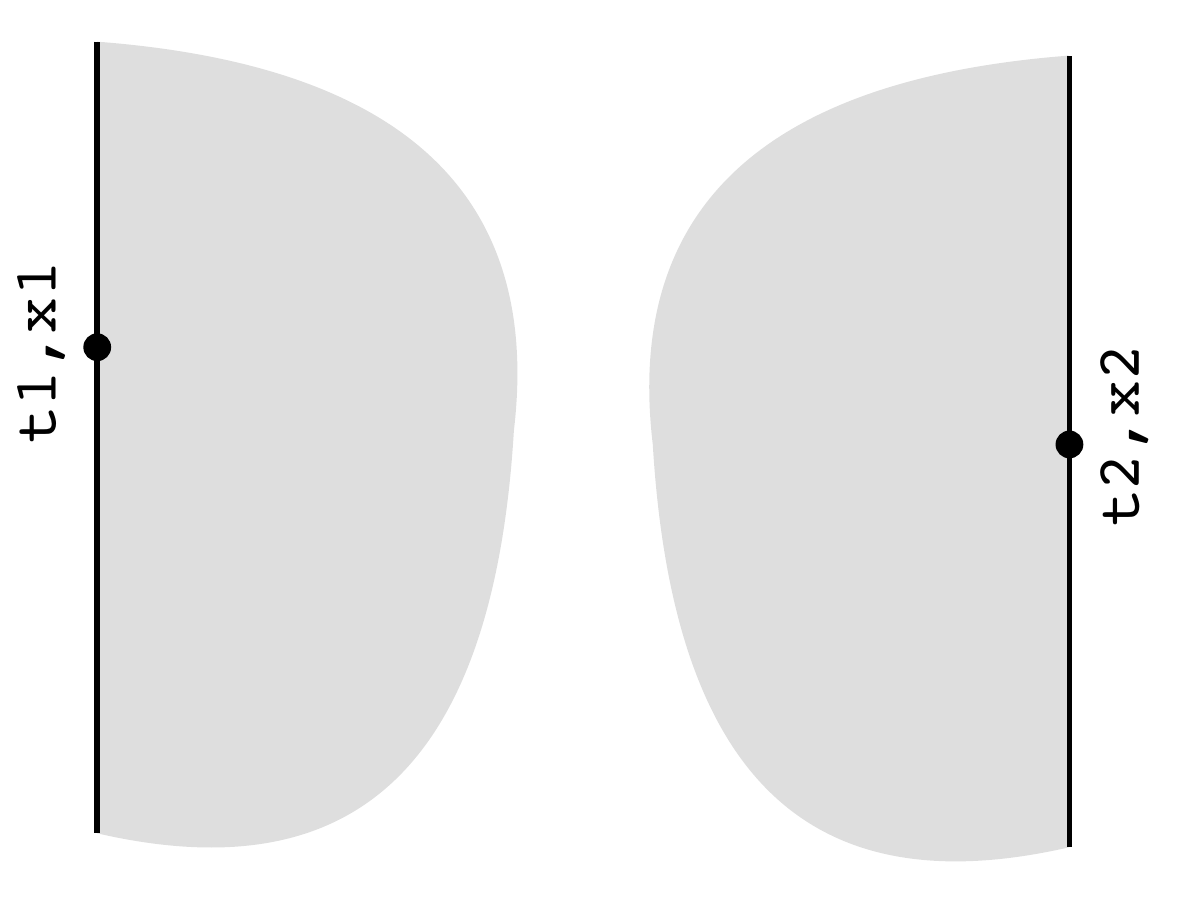} 
	\begin{picture}(0.1,0.25)(0,0)
\put(-105,162){\makebox(0,0){$|\Psi\rangle = |\Psi_A\rangle \otimes |\Psi_B\rangle$}}
\put(-200,18){\makebox(0,0){$A$}}
\put(-10,18){\makebox(0,0){$B$}}
\end{picture}
	\hskip2em
	\includegraphics[width=7cm]{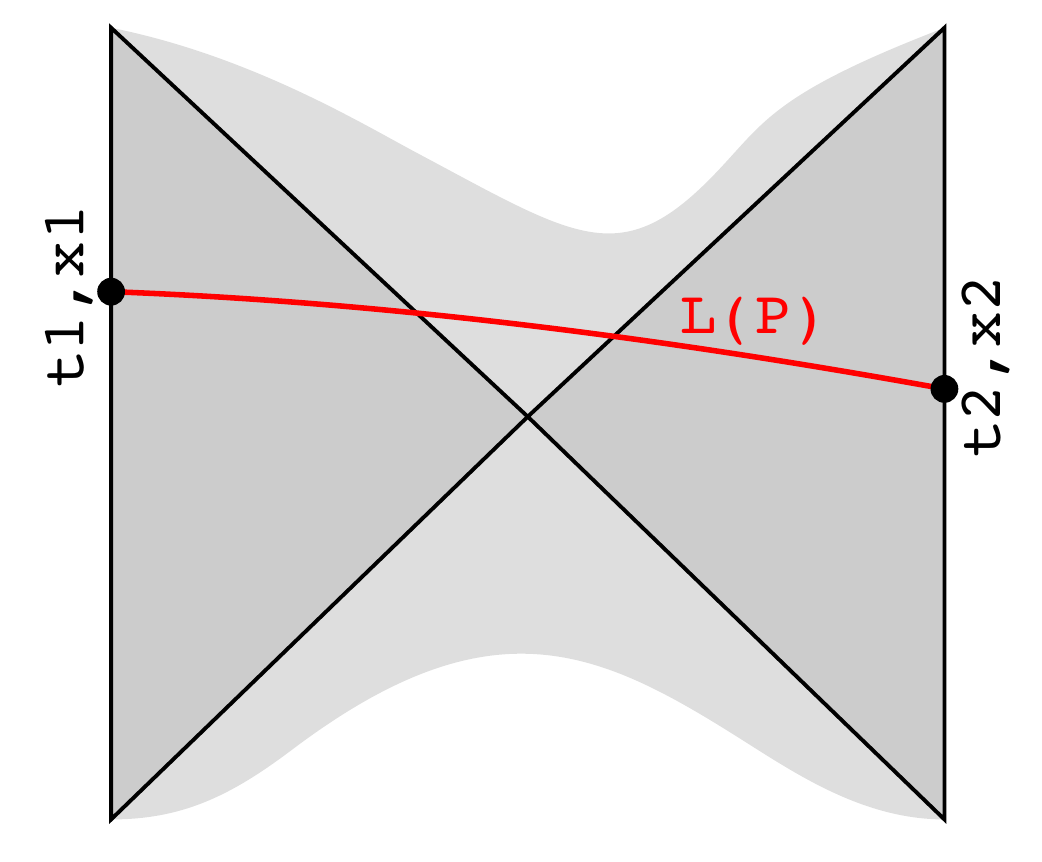}
	\begin{picture}(0.1,0.25)(0,0)
	\put(-195,18){\makebox(0,0){$A$}}
\put(-10,18){\makebox(0,0){$B$}}
\put(-105,162){\makebox(0,0){$|\Psi\rangle \neq |\Psi_A\rangle \otimes |\Psi_B\rangle$}}
\end{picture}
\end{center}
\caption{
	\label{F:geodesic}
	\textbf{(Left)} A cartoon of the gravity dual for a CFT on two disconnected regions $A$ and $B$, where the theory is in a product state. There are no bulk trajectories which go from one boundary to the other. \textbf{(Right)} The gravity dual of an entangled state. The dual spacetime is connected, and there are paths from $A$ to $B$ through the bulk. The diagonal black lines are event horizons, so that there are no timelike paths from $A$ to $B$. This is an ER bridge.
}
\end{figure}
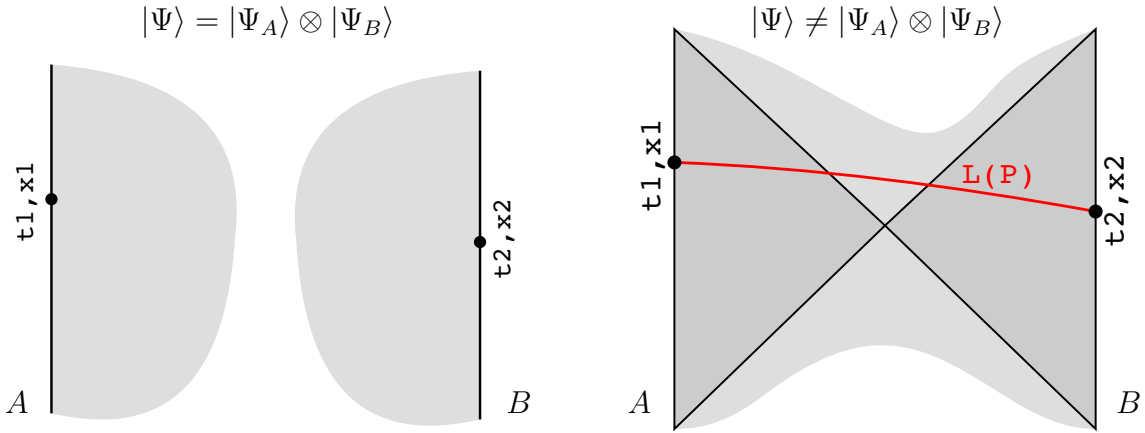

To proceed, we find it useful to revisit the connectedness of the dual spacetime~\cite{VanRaamsdonk:2010pw} by studying correlators of local operators as in~\eqref{E:twoPoint}. In the $N\gg \lambda\gg 1$ limits, a two-point function of dimension-$\Delta$ operators is given by a path integral over bulk curves ${\cal P}(1,2)$ connecting the insertions, which are on the AdS boundary~\cite{Louko:2000tp},
\begin{equation}
\label{E:geodesic}
\langle {\cal O}(t_1,\mathbf{x}_1)\,{\cal O}(t_2,\mathbf{x}_2)\rangle = \int {\cal DP}(1,2)\, e^{i m  L(\cal P)} \sim e^{i \Delta L_{\rm geo.}(1,2)}\,.
\end{equation}
Here, timelike paths have real length. In the large $\Delta$ limit (with $\Delta(\Delta-4)=m^2$), the integral is approximately equal to its stationary phase approximation, which is the exponential of ($i\Delta$ times) the length of the shortest bulk geodesic connecting the insertions. So nonzero spacelike correlations imply the existence of bulk geodesics connecting spatially separated regions on the boundary, i.e. those correlations imply spatial bulk connectedness, and vice versa. See Fig.~\ref{F:geodesic}. Moreover, those correlations decrease as $L_{\rm geo}$ increases. 

Consider now two causally disconnected sets of degrees of freedom, $A$ and $B$, and suppose each has a gravity dual. Then the bulk has two asymptotically AdS regions with boundaries $A$ and $B$. If $A$ and $B$ are entangled, then there are nonzero correlations between operators in $A$ with those in $B$, which by~\eqref{E:geodesic} implies that there is a finite-length spacelike curve connecting the two boundaries. On the other hand, causality implies that no timelike curves connect the boundaries. We must therefore have bulk horizons that shield one asymptotic region from the other. From these simple arguments, we see that the gravity dual of EPR entanglement is a connected spacetime between different asymptotically AdS regions with horizons between them. This is a non-traversable wormhole, or Einstein-Rosen (ER) bridge. 

Note that this argument only tells us that such a wormhole exists, and says nothing about whether it is smooth or weakly curved behind the horizons. We cannot expect that the geometry dual to a general entangled state is everywhere smooth and weakly curved~\cite{Marolf:2013dba,Balasubramanian:2014gla}.

The simplest example of such a wormhole in AdS/CFT is the dual of a tensor-product CFT with CFTs on $k$ different spatial spheres. In an unentangled state, the dual is $k$ disconnected copies of global AdS, but the dual of an entangled state is a connected spacetime with $k$ asymptotic AdS regions, shielded from each other by a $k$-sided horizon. When $k=2$, the eternal AdS black hole is an example, dual to the thermofield double state of the CFT~\cite{Maldacena:2001kr}. For $k>2$ and $d=2$ CFT, these wormholes have been constructed in~\cite{Skenderis:2009ju}.

It was essential for our argument that $A$ and $B$ had $\mathcal{O}(N^2)$ degrees of freedom, ``large'' enough to have a gravity dual. However, there are entangled states that do not give rise to a wormhole geometry, for example by adding $\mathcal{O}(1)$ entangled excitations on top of the CFT vacuum, or cases where boundary causality does not require bulk horizons~\cite{Nozaki:2013wia,Asplund:2013zba}. The dual of a pair of entangled quarks is a fundamental string which contributes ${\cal O}(\sqrt{\lambda})$ entangled excitations, so na\"ively it doesn't have a wormhole either. However, there is an analogous argument that the dual of entangled quarks contains a wormhole.

\section*{Holographic EPR pairs}

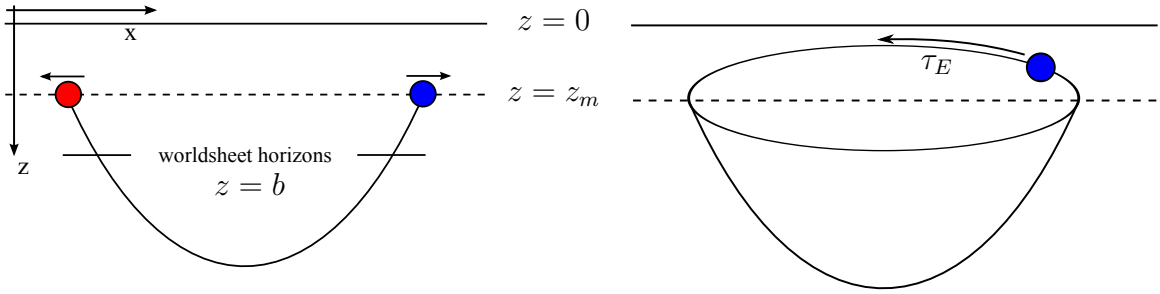
\begin{figure}[t]
\begin{center}
	\includegraphics[width=7cm]{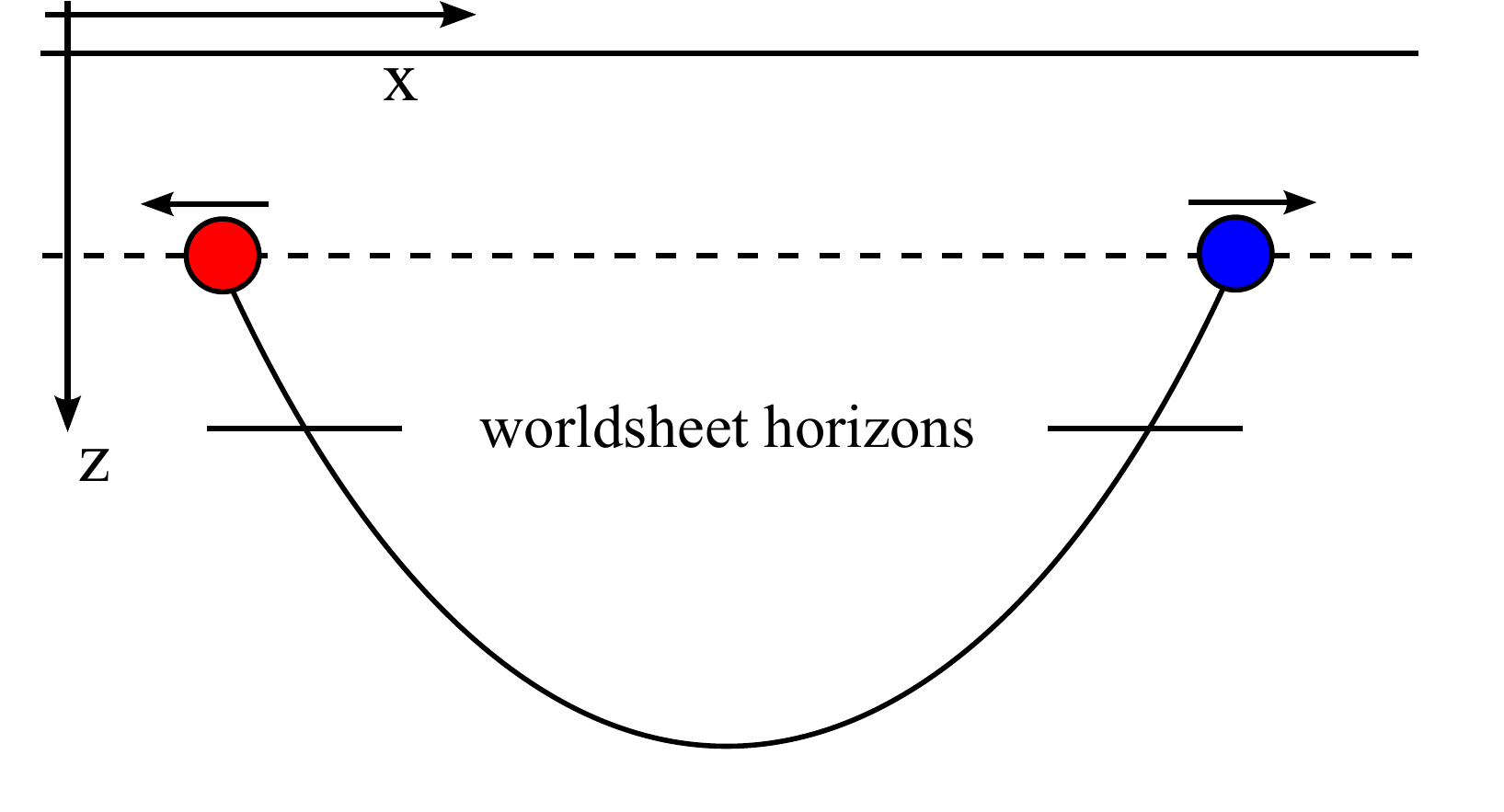}
	\begin{picture}(0.1,0.25)(0,0)
\put(10,102){\makebox(0,0){$z = 0$}}
\put(10,73){\makebox(0,0){$z = z_m$}}
\put(-105, 40){\makebox(0,0){$z = b$}}
\put(155,85){\makebox(0,0){$\tau_E$}}
\end{picture}
	\hskip3em
	\includegraphics[width=7cm]{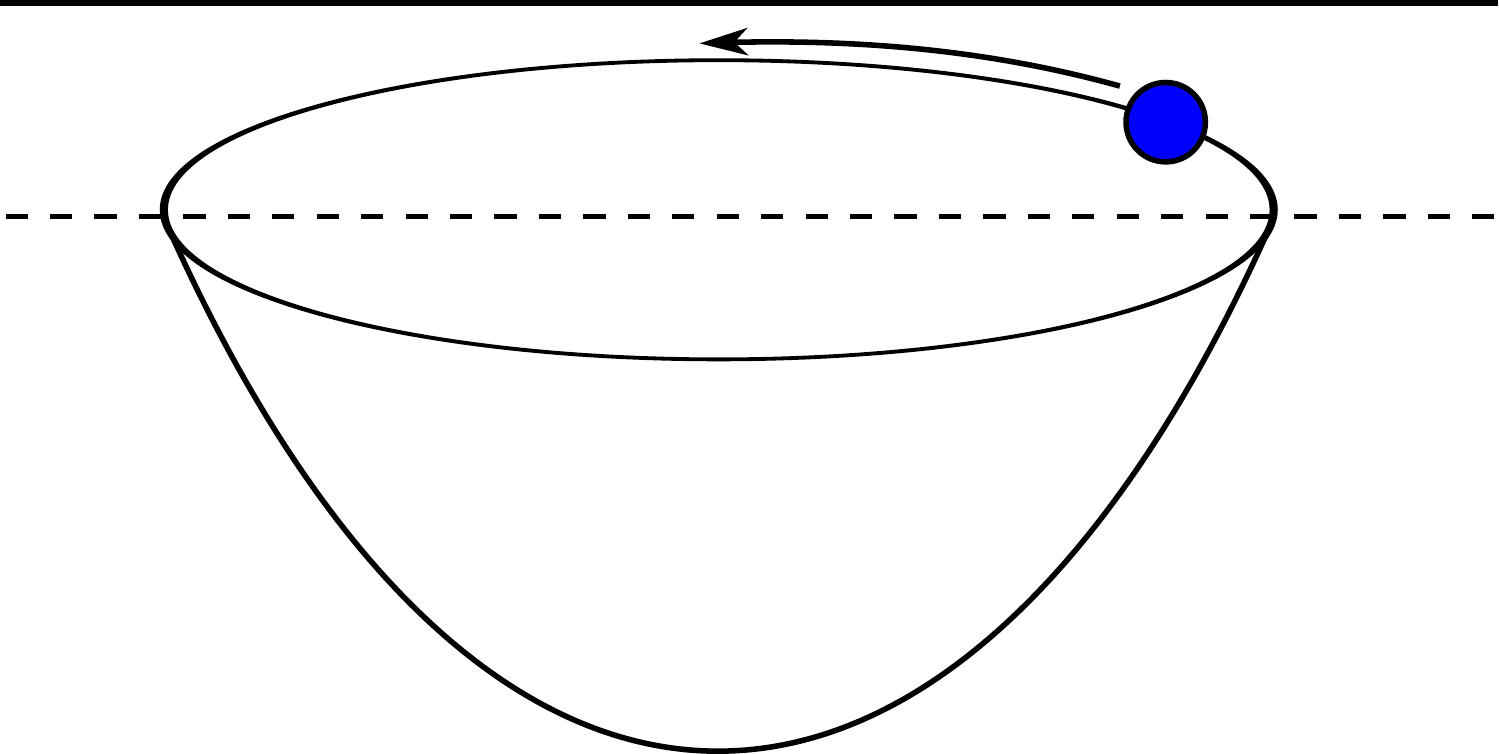}
\end{center}
\caption{
	\label{F:qqbar} \textbf{(Left)} The expanding string dual to the $q$-$\bar{q}$ pair, shown at a moment in time. \textbf{(Right)} The Euclidean worldsheet instanton which describes the creation of the $q$-$\bar{q}$ pair in terms of a quark circulating in Euclidean Rindler time $\tau_E$.} 
\end{figure}

Consider a quark-antiquark ($q$-$\bar{q}$) pair added to $\mathcal{N}=4$ SYM, entangled into a color singlet and undergoing constant acceleration. As far as we know, this is the closest we can get to an EPR pair in holography \cite{Jensen:2013ora,Sonner:2013mba} (see also~\cite{Chernicoff:2013iga}). The dual is a single fundamental string whose endpoints represent the quarks. For quarks of mass $m$, the string ends on a flavor brane~\cite{Karch:2002sh} at $z=z_m = \sqrt{\lambda}/(2\pi m)$, in coordinates where the AdS metric is
\begin{equation*}
g = \frac{dz^2 - dt^2 + d\vec{x}^2}{z^2}\,.
\end{equation*}
The AdS boundary is at $z=0$. Turning on an electric field $E=m/b$ on the brane, entangled $q$-$\bar{q}$ pairs are created via the Schwinger effect and accelerated away from each other. The dual string is an expanding semi-circle~\cite{Xiao:2008nr}
\beq
\label{E:theString}
x^2 = t^2 - z^2 + b^2\,.
\eeq
The quarks uniformly accelerate along $x_{\pm} =\pm \sqrt{t^2+a^2}$ with $a^2=b^2-z_m^2$. The whole process is described by patching together a Euclidean instanton describing the string creation with the Lorentzian solution~\eqref{E:theString} at $t=0$, as depicted in Fig.~\ref{F:qqbar}. 

\begin{figure}[t]
\begin{center}
	\includegraphics[width=8cm]{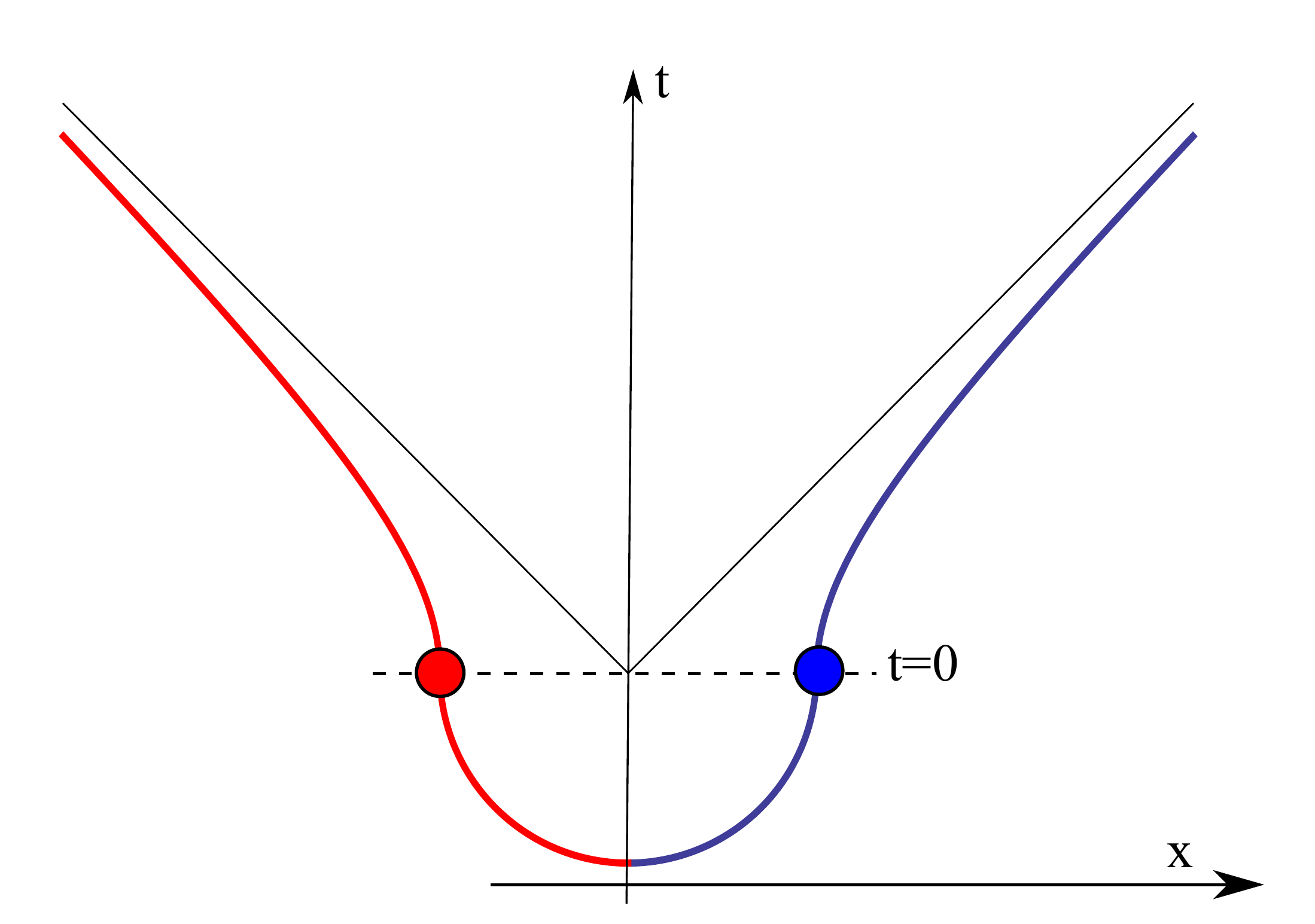}
\end{center}
\caption{
	\label{F:euclideanWS} The patching of the Euclidean instanton describing the pair creation with the Lorentzian evolution of the $q$-$\bar{q}$ pair. The colored lines indicate the trajectories of the quark/anti-quark.}
\end{figure}

\begin{figure}[t]
\begin{center}
	\includegraphics[width=7cm]{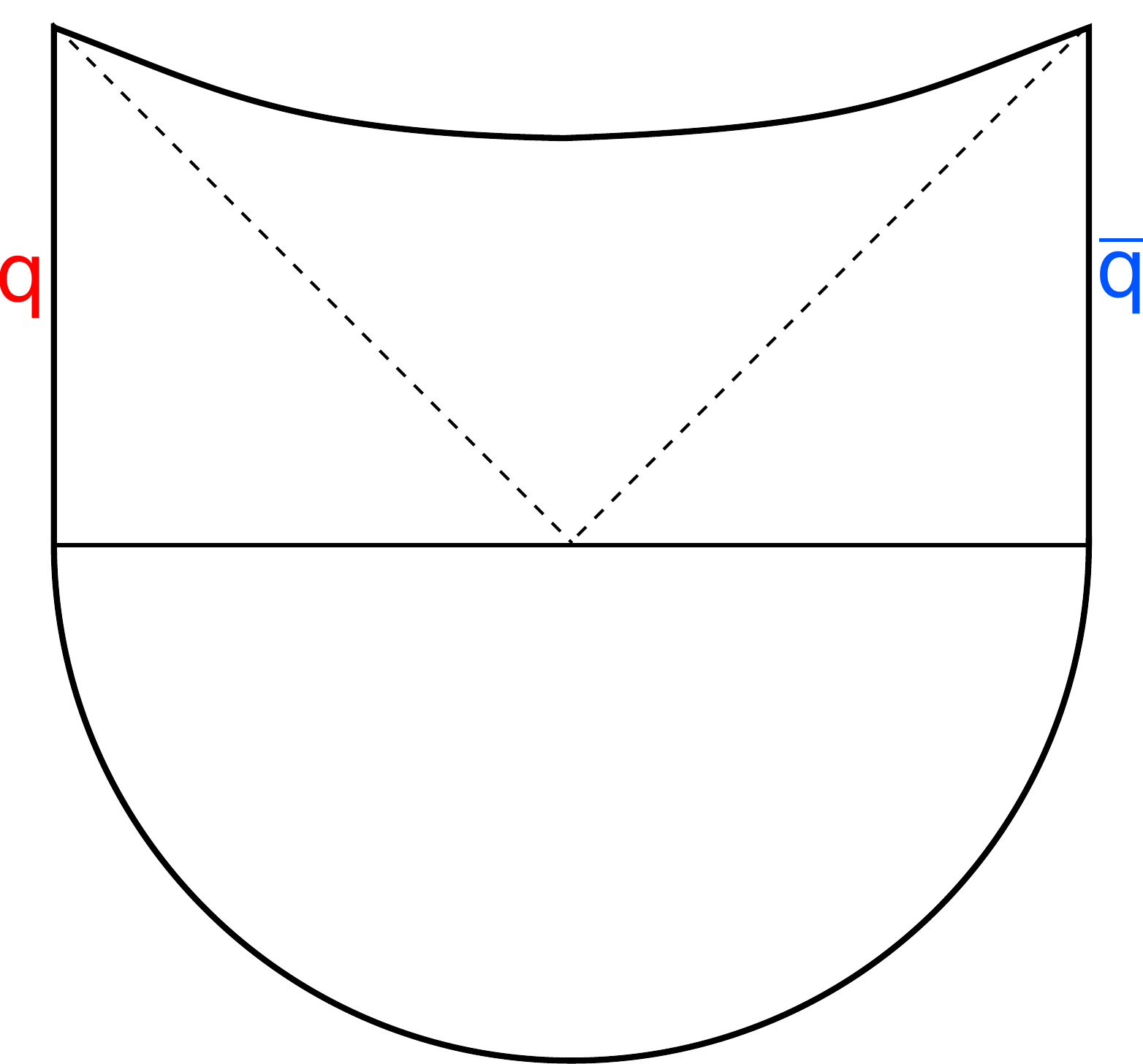}
\end{center}
\caption{
	\label{F:penrose} The worldsheet Penrose diagram for the Hartle-Hawking construction of the entangled $q$-$\bar{q}$ pair.
}
\end{figure}

This setup exhibits EPR entanglement between the quarks since they are entangled into a color singlet, and are out of causal contact for all time. Indeed, the Euclidean/Lorentzian patching tells us that the state of the pair is, in close analogy with the eternal black hole~\cite{Maldacena:2001kr},
\beq
\label{E:qqbarPsi}
|\Psi\> = \sum_{n,c,\bar{c}} e^{- \beta E_n/2} \delta^{c\bar{c}}|n,c\>_q\otimes |n,\bar{c}\>_{\bar{q}}\,,
\eeq
where
\beq
\beta = 2\pi a
\eeq
is the inverse Rindler temperature. The Euclidean/Lorentzian patching can be viewed as the Hartle-Hawking construction for the $q$-$\bar{q}$ state. The sum~\eqref{E:qqbarPsi} runs over single-quark eigenstates of the Rindler Hamiltonian and identifies their color $c$, so $|\Psi\>$ is gauge-invariant. There are connected correlators which quantify the $q$-$\bar{q}$ entanglement, given by an expression like~\eqref{E:geodesic} in the large $\lambda$ limit, where now the operators act on the quark and anti-quark Hilbert spaces. The curves $\mathcal{P}$ in~\eqref{E:geodesic} become curves on the string worldsheet. Running through the same arguments as above, {\it mutatis mutandis}, we conclude that EPR entanglement between the quarks implies that the induced metric of the dual string has a wormhole. Indeed, changing coordinates as 
\begin{equation*}
x\pm t = b\sqrt{1-\frac{w^2}{a^2}} e^{\pm \frac{\tau}{a}}\,, \qquad z = \frac{b w}{a}\,,
\end{equation*}
the worldsheet metric is the $AdS_2$ ``black hole''\footnote{In the same sense as in~\cite{Spradlin:1999bn}.} 
\beq
\label{E:WSwormhole}
\gamma=\text{P}[g] =\frac{1}{w^2} \left[-\left( 1-\frac{w^2}{a^2}\right) d\tau^2 + \frac{dw^2}{1-\frac{w^2}{a^2}} \right]\,.
\eeq
Its conformal diagram, Fig.~\ref{F:penrose}, illustrates the Hartle-Hawking construction of the eternal AdS$_2$ black hole~\cite{Spradlin:1999bn,Maldacena:2001kr}. The two asymptotic regions are near the quark and antiquark. Furthermore, using the CHM map~\cite{Casini:2011kv,Jensen:2013ora,Jensen:2013lxa}, the $q$-$\bar q$ pair contributes to the entanglement entropy of a Rindler wedge as
\beq
\label{E:EE}
S_{EE} = S_{\mathcal{N}=4} + \frac{\sqrt{\lambda}}{3} + \mathcal{O}(\ln(N),\ln(\lambda))\,.
\eeq

This setup survives essentially unchanged under the addition of dynamical gravity to $\mathcal{N}=4$ through the Randall-Sundrum~\cite{Randall:1999vf} scenario~\cite{Jensen:2014bpa}. One can also prepare an initial state which is~\eqref{E:qqbarPsi} plus fluctuations, by adding fluctuations of the string during the Euclidean part of its evolution. The worldsheet will still have a wormhole. Finally, worldvolume horizons have shown up in holography before; see e.g.~\cite{Gubser:2006nz}; their physical interpretation has been somewhat unclear. This example suggests that worldvolume horizons indicate EPR entanglement between colored degrees of freedom.

\section*{Discussion}

This work was inspired by the ER-EPR conjecture~\cite{Maldacena:2013xja}. We have clarified what the conjecture means in holography. First, not all entanglement patterns in CFT are geometrized into wormholes. Rather (but still remarkably), the holographic dual of EPR entanglement is a wormhole. The basic rules of AdS/CFT tell us that such a wormhole must exist, but the dual of a general entangled state may not be everywhere smooth and weakly curved. Second, the dual wormholes provide a good description of entanglement when there is a good notion of spacetime (and worldsheet) geometry. In the small $N$ and $\lambda$ limits, CFT entanglement is better described with field theory correlations like~\eqref{E:twoPoint}, which may be taken to define a ``stringy wormhole.'' Third, for CFT regions, which are entangled and remain forever out of causal contact, the gravity dual contains a wormhole; one requires $\mathcal{O}(N^2)$ entanglement to have a wormhole in the dual spacetime, and $\mathcal{O}(\sqrt{\lambda})$ entanglement for a wormhole on the induced geometry of a dual string. 

We conclude with some interesting open questions. What is the holographic dual of decoherence, after bringing EPR entangled CFTs into causal contact? Are there wormholes in holography that bridge causally connected exteriors, and if so, what is the corresponding entry in the AdS/CFT dictionary? Finally, can one use exact results~\cite{Lewkowycz:2013laa} for e.g.~\eqref{E:EE} to infer what happens to the worldsheet wormhole~\eqref{E:WSwormhole} at finite $N$ and $\lambda$?

\section*{Acknowledgements}

We thank Mike Crossley, Andreas Karch, Hong Liu, and Josephine Suh for discussion and related collaboration. This work was supported by the National Science Foundation under grant PHY-0969739, as well as the Department of Energy under cooperative research agreement Contract Number DE-FG02-05ER41360.

\bibliography{refs}
\bibliographystyle{JHEP}

\end{document}